\documentstyle[epsfig,preprint,aps]{revtex}
\begin{document}
\draft
\title{Self-similar collapse of a massless scalar field in
three-dimensions.} 
\author{G. Oliveira-Neto\thanks{Email:
gilneto@fisica.ufjf.br}} 
\address{Departamento de F\'{\i}sica,
Instituto de Ciencias Exatas,
Universidade Federal de Juiz de Fora,
CEP 36036-330, Juiz de Fora,
Minas Gerais, Brazil.}
\date{\today}
\maketitle

\begin{abstract}
We study an analytical solution to the Einstein's equations in
$2+1$-dimensions, representing the self-similar collapse of a
circularly symmetric, minimally coupled, massless, scalar field.
Depending on the value of certain parameters, this solution
represents the formation of black holes. Since our solution is
asymptotically flat, our black holes do not have the BTZ 
space-time as their long time limit. They represent a new family
of black holes in $2+1$-dimensions.
\end{abstract}

Since the work of M. W. Choptuik on the gravitational collapse 
of a massless scalar field Ref. \cite{choptuik}, many physicists 
have focused their attentions on the issue of gravitational 
collapse. In that work Choptuik showed that one may understand 
the gravitational collapse as a critical phenomena. In a sense 
that there is a critical value for a certain parameter, in the
parameter space of solutions, which separates two types of 
solutions: those which are black holes from others which are not
black holes.

An important arena where one can study the gravitational collapse
is general relativity in $2+1$-dimensions. The great appeal of
this theory comes from the fact that it retains many of the
properties of general relativity in $3+1$-dimensions, but the
field equations are greatly simplified \cite{jackiw}.

Presently, several black hole solutions in $2+1$-dimensional
general relativity are known \cite{mann}. Including the
first one to be discovered, the so-called BTZ black hole
\cite{banados}. All of them have an important property in
common: the presence of a negative cosmological constant,
which makes them asymptotically anti-de Sitter. 

Indeed, in a recent work it was demonstrated that a 
three-dimensional solution to the Einstein's equations,
with a positive cosmological constant ($\Lambda$), such that
the stress-energy tensor satisfies the dominant energy
condition, contains no apparent horizons \cite{ida}. The same
result applies to the case $\Lambda = 0$ in the presence of
matter fields. Therefore, this result explains the necessity
of a negative cosmological constant in order to a black hole
to form, in three-dimensional general relativity.

Considering the conditions used to derive the, above mentioned,
theorem it might be possible to have the formation of an
apparent horizon in a space-time in three-dimensions, without
a negative cosmological constant. For this, the matter content
of this space-time should have a stress-tensor that does not
satisfy the dominant energy condition. It is this possibility
that we shall investigate here. We shall allow the, initially 
real, scalar field to become complex which will force its
stress-energy tensor to violate the dominant energy condition.
We are aware that there is a great debate whether this matter
content is physically acceptable or not \cite{thorne}, but the
theoretical possibility of a new family of black holes in
$2+1$-dimensions, we believe, is enough motivation to use it 
here.

In the present letter we would like to present a solution to the
Einstein's equation, without a cosmological constant, representing
the self-similar, circularly symmetric, collapse of a minimally
coupled, massless, scalar field, in $2+1$-dimensions. As we shall
see this solution, depending on the value of certain parameters, 
represents the formation of black holes as the result of the
collapse process.

We shall start by writing down the ansatz for the space-time metric.
As we have mentioned before, we would like to consider the circularly
symmetric, self-similar, collapse of a massless scalar field in 
$2+1$-dimensions. Therefore, we shall write our metric ansatz as,

\begin{equation}
\label{1}
ds^2\, =\, -\, 2 e^{2\sigma(u,v)} du dv\, +\, r^2(u,v) d\theta^2
\, ,
\end{equation}
where $\sigma(u,v)$ and $r(u,v)$ are two arbitrary functions to be
determined by the field equations, $(u,v)$ is a pair of null
coordinates varying in the range $(-\infty,\infty)$, and $\theta$
is an angular coordinate taking values in the usual domain 
$[0,2\pi]$.

The scalar field $\Phi$ will be a function only of the two null
coordinates and the expression for its stress-energy tensor
$T_{\alpha\beta}$ is given by \cite{wheeler},

\begin{equation}
\label{2}
T_{\alpha\beta}\, =\, \Phi,_\alpha \Phi,_\beta\, -\,
{1\over 2} g_{\alpha\beta} \Phi,_\lambda \Phi^{,_\lambda}\, .
\end{equation}
where $,$ denotes partial differentiation.

Now, combining Eqs. (\ref{1}) and (\ref{2}) we may obtain the
Einstein's equations which in the units of Ref. \cite{wheeler}
and after re-scaling the scalar field, so that it absorbs the 
appropriate numerical factor, take the following form,

\begin{equation}
\label{3}
2\sigma,_u r,_u\, -\, r,_{uu}\, =\, r (\Phi,_u)^2\, ,
\end{equation}
\begin{equation}
\label{4}
2 \sigma,_v r,_v\, -\, r,_{vv}\, =\, r (\Phi,_v)^2\, ,
\end{equation}
\begin{equation}
\label{5}
2 r \sigma,_{uv}\, +\, r_{uv}\, =\, -\, r (\Phi,_u
\Phi,_v)\, ,
\end {equation}
\begin{equation}
\label{6}
r,_{uv}\, =\, 0\, ,
\end{equation}
The equation of motion for the scalar field, in these 
coordinates, is

\begin{equation}
\label{7}
2 r \Phi,_{uv}\, +\, \Phi,_{v} r,_u\, +\, \Phi,_{u} r,_v\,
=\, 0\, .
\end{equation}

The above system of non-linear, second-order, coupled, partial
differential equations (\ref{3})-(\ref{7}) has an analytical
solution if we impose that it is continuously self-similar. More
precisely, following Coley \cite{coley}, our system will have
self-similarities of the first and second kinds.

Under these conditions our solution will be given by,

\begin{equation}
\label{8}
r(u,v)\, =\, \beta (\alpha v)^{1/\alpha}\, +\, \gamma u\, ,
\end{equation}
\begin{equation}
\label{9}
\sigma (u,v)\, =\, \left({1 - \alpha\over 2}\right)\ln{\left({r
\over u}\right)}\, +\, \sigma_0\, ,
\end{equation}
and the scalar field has the following value,
\begin{equation}
\label{10}
\Phi (u,v)\, =\, (1 - \alpha)^{1/2} \ln \left[{ 
\sqrt{(\gamma/\beta)u}\, -\, \imath 
\sqrt{(\alpha v)^{1/\alpha}}\over 
\sqrt{(\gamma/\beta)u}\, +\, \imath 
\sqrt{(\alpha v)^{1/\alpha}}}\right]\,,
\end{equation}
where $\gamma$ and $\beta$ are real, integration constants and
$\alpha$ is a positive real number associated with the kinematic,
continuous, self-similarity. For $\alpha=1$, the self-similarity
is of the first kind, for $0 < \alpha < 1$ the self-similarity is
of the second kind \cite{coley}. Following \cite{brady}, we shall
assume that $\Phi(u,v) \equiv 0$ for $v < 0$.

In terms of $r(u,v)$ Eq. (\ref{8}), and $\sigma(u,v)$ Eq. (\ref{9}),
the line element Eq. (\ref{1}) becomes,

\begin{equation}
\label{12}
ds^2\, =\, - 2 e^{2\sigma_0} \left({r\over u}\right)^{(1-\alpha)}
du dv\, +\, r^2 d\theta^2\, .
\end{equation}
One may notice from Eqs. (\ref{8}-\ref{10}), that for different 
values of $\alpha$, $\beta$ and $\gamma$, one has different 
space-times.

Observing Eq. (\ref{12}), we notice that these space-times have a
singularity at $r=0$. It is a physical singularity as can be seen
directly from the curvature scalar $R$.

In order to show this result we start writing down the Ricci tensor
that, in the present case, has the following expression \cite{taub},

\begin{equation}
\label{13}
R_{\alpha \beta}\, =\, \Phi,_\alpha\, \Phi,_\beta \, .
\end{equation}
From it, we may compute $R$ straightforwardly with the aid of
Eqs. (\ref{8})-(\ref{12}), finding,

\begin{equation}
\label{14}
R\, =\, - 2 (1 - \alpha) \gamma \beta e^{-2\sigma_0}{[(\alpha 
v)^{1/\alpha} u]^{(1-\alpha)}\over r^{(3-\alpha)}}\, .
\end{equation}

Finally, taking the limit $r \to 0$ in $R$ Eq. (\ref{14}), we
find that this quantity diverges at $r=0$. There is no other physical
singularity for these space-times because $R$ is well defined outside
$r=0$. In particular, $u=0$ is just an apparent singularity and a new
coordinate system can be found where it disappears. As we shall see
below, $u=0$ is an apparent horizon for these space-times. 

Another important property we can learn from $R$ is the asymptotic
behavior of our solution. If we take the limit $r \to \infty$ of $R$
Eq. (\ref{14}), we find that $R \to 0$. Therefore, we conclude that 
the space-times under investigation are asymptotically flat.

The apparent horizons are determined by imposing that
the surface $r$ = constant becomes null, which implies that,

\begin{equation}
\label{25}
2g^{uv}\, r,_u \, r,_v\, =\, 0\, .
\end{equation}
For the above space-times, this equation (\ref{25}) takes the
following form when we introduce the appropriate information from
Eqs. (\ref{8}) and (\ref{12}),

\begin{equation}
\label{26}
- 2 e^{-2\sigma_0} \gamma \beta \left[ {(\alpha v)^{1/\alpha} u
\over r}\right]^{(1-\alpha)}\, =\, 0\, .
\end{equation}

In the most general case, the space-times under study may have
two distinct apparent horizons, from the solutions of Eq. 
(\ref{26}). They are the surfaces $u = 0$ and $v = 0$.

The space-times above
will only be physically relevant for the collapse process if 
$r$ Eq. (\ref{8}) is a real, positive function. Also, at
least outside the horizon, $r$ = constant, must be a set of time-like 
surfaces for different constants. On the other hand, as we have 
mentioned above the scalar field should be allowed to take 
complex values if we want to obtain black hole solutions.

It is clear from Eqs. (\ref{12}) and (\ref{14}) that $\alpha=1$
is the three-dimensional Minkowski space-time. Therefore, we
shall restrict our attention to the space-times with
self-similarity of the second kind. From Eq. (\ref{10}), it is
appropriate to consider the influx of scalar field to be
turned on at the advanced time $v=0$. So that to the past of
this surface the space-time is Minkowskian and the metric is
therefore $C^1$ at this surface.

The black hole space-times are obtained for $\gamma < 0$ and
$\beta > 0$. Figure $1$ shows a conformal diagram for a 
typical space-time in this case. We may see that the space-time
is divided, naturally, in three distinct regions. The first
one is the Minkowskian region where $v < 0$ (I). Then, we
have the external region where $v >0$ and $u < 0$ (II).
Finally, in the internal region $u > 0$ (III).

The scalar field starts collapsing from past null infinity,
in the external region. From eq. (\ref{10}), it is not
difficult to see that it is imaginary in this region. It is
zero at $v=0$, grows to $\Phi = \imath (\pi/2) 
\sqrt{1-\alpha}$ at $u = (\beta/\gamma)(\alpha v)^{1/\alpha}$
and reaches its maximum value, in this region, $\Phi = \imath
\pi \sqrt{1-\alpha}$, at the apparent horizon $u=0$. In the
internal region we may re-write $\Phi(u,v)$ Eq. (\ref{10}) as,

\begin{equation}
\label{27}
\Phi(u,v) = \sqrt{1-\alpha} [ \ln{r}\, -\, 
2\ln{(\sqrt{|\gamma|u} + \sqrt{\beta (\alpha v)^{1/\alpha}})}
 + \imath\pi ]\, .
\end{equation}
From it we see that $\Phi(u,v)$ becomes complex with a 
constant imaginary part. Its real part decreases from zero 
at the horizon $u=0$, until it blows up at the singularity 
$r=0$.

Observing Eqs.(\ref{25}) and (\ref{26}), we see that the
surfaces $r=$constant will be time-like in both external
and internal regions if $1-\alpha=l/m$, where $l$ and
$m$ are integer numbers, $l$ being even and $m$ being odd.
Therefore, for this choice of $\alpha$, the singularity 
$r=0$ will be a time-like one. Once that this singularity
is hidden by the horizon $u=0$, we may consider these 
space-times representing black holes.

Since our solutions are asymptotically flat, our black 
holes do not have the BTZ space-time as their long time
limit. They represent a new family of black holes in
$2+1$-dimensions.

We may also describe our solution with the aid of the 
time coordinate,

\begin{equation}
\label{28}
t(u,v)\, =\, \beta (\alpha v)^{1/\alpha}\, -\, \gamma u
\, ,
\end{equation}
in terms of which, the line element Eq. (\ref{12}) and the
scalar field Eq. (\ref{10}) become, respectively,

\begin{equation}
\label{29}
ds^2 = -{(2)^{(1-2\alpha)}e^{2\sigma_0}\over (\beta
\gamma)^{\alpha}}\left({r\over r^2 -t^2}\right)^{(1-\alpha)}
(- dt^2 + dr^2) + r^2 d\theta^2
\end{equation}
and
\begin{equation}
\label{30}
\Phi(r,t)\, =\, \sqrt{1-\alpha}\, \ln \left[\, {\sqrt{r-t} - 
\imath \sqrt{r+t}\over \sqrt{r-t} + \imath \sqrt{r+t}}\,
\right]\, .
\end{equation}

We end the letter by noting that recently another
continuously self-similar solution, to the same problem
treated here, was found \cite{garfinkle}. It is not
difficult to realize that the solution described here is
different from the one derived in \cite{garfinkle}, because
the latter has a self-similarity of the first kind 
\cite{coley}. As we have seen, if we set the condition
that our solution has just a self-similarity of the first
kind ($\alpha = 1$), it reduces to Minkowski space-time.
Therefore, they are not the same solution.

I am grateful to A. Wang for suggestive discussions 
in the course of this work and FAPEMIG for the invaluable
financial support.

\begin{figure}
\psfig{file=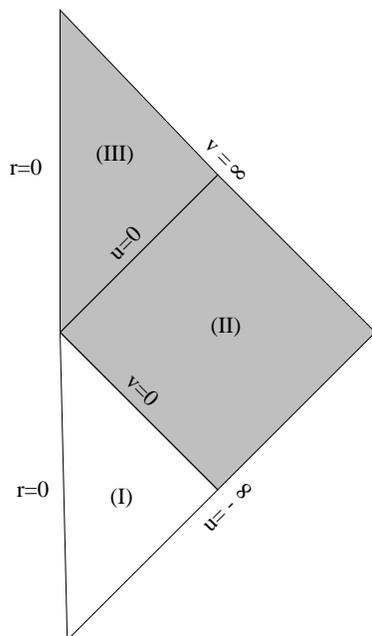}
\caption{Conformal diagram for a typical black hole 
solution. The apparent horizon $u=0$ separates the
Minkowskian (I) and external (II) regions from the 
internal region (III) where lies the time-like 
singularity at $r=0$.}
\end{figure}

\end{document}